\documentclass[a4paper,aps,10pt]{revtex4}
\usepackage{amsmath}
\usepackage{graphics}
\usepackage{mathrsfs}

\begin{document}

\title{Liquid-glass transition of a fluid confined in a disordered
porous matrix: A mode coupling theory}

\author{V. Krakoviack}

\affiliation{Laboratoire de Chimie, \'Ecole Normale Sup\'erieure de
Lyon, 46, All\'ee d'Italie, 69364 Lyon Cedex 07, France}

\date{\today}

\begin{abstract}
We derive an extension of the mode coupling theory for the
liquid-glass transition to a class of models of confined fluids, where
the fluid particles evolve in a disordered array of interaction
sites. We find that the corresponding equations are similar to those
describing the bulk, implying that the methods of investigation which
were developed there are directly transferable to this new domain of
application. We then compute the dynamical phase diagram of a simple
model system and show that new and nontrivial transition scenarios,
including reentrant glass transitions and higher-order singularities,
can be predicted from the proposed theory.
\end{abstract}

\maketitle

Since the mode coupling theory (MCT) of the liquid-glass transition
was proposed in the mid-eighties \cite{leu84pra,bengotsjo84jpc}, it
has acquired a central role in this field of research
\cite{leshouches,gotsjo92rpp,got99jpcm}. Indeed, on the experimental
or numerical side, it is very often to the predictions of the MCT that
new data are first confronted \cite{gotsjo92rpp,got99jpcm}, and on the
theoretical side, models of increasing complexity are regularly
investigated within the MCT framework as a means to unveil potential
new phenomena \cite{highorder}. The reason for this strong influence
of the MCT lies in its ability to reproduce important phenomenological
aspects of the dynamics of supercooled liquids: First, of course, the
slowing down of the structural relaxation when density is increased or
temperature decreased, but other more specific features as well, like
the two-step relaxation scenario. Moreover, it makes a number of
precise universal predictions especially suitable for comparisons with
experiment or simulation results, and, for simple enough systems for
which the MCT is tractable as a first principle theory, it provides
detailed predictions concerning nonuniversal aspects of the dynamics
as well, allowing thus extensive quantitative tests of the theory
\cite{gotsjo92rpp,got99jpcm}.

In the past few years, a rapidly growing interest for the dynamics of
liquids under confinement has built into the glass transition
community \cite{proceedings}, with the aim of clarifying the concept
of cooperativity, a key ingredient of many glass transition theories
\cite{heterogeneities}. Indeed, confinement has appeared as a means to
impose to a glassforming system a new characteristic lengthscale (pore
size, film thickness\dots), which should interact with any correlation
or cooperativity length developing in it, possibly leading to indirect
informations on the nature and evolution with temperature of
cooperativity. In the course of these investigations, at least for
some systems studied by computer simulation
\cite{galrovspo00prl,galpelrov02el,schkolbin04jpcb}, many features of
the dynamics of bulk glassforming liquids which had found an
interpretation in the framework of the MCT have been uncovered. It
then appeared natural to compare the corresponding data with the
predictions of the MCT, even if the theory had been designed for bulk
fluids, and, because the tests were quite successful, the idea emerged
that a mode coupling scenario was at work in confined fluids as well.

It seems thus that there is a clear need for an extension of the MCT
to confined glassforming liquids. First, if this theory turned out to
be similar enough to the theory for the bulk, this would put the
studies of simulation data mentioned above on firmer grounds, and more
generally this would provide a framework for the analysis of
experimental or numerical data on confined fluids. Second, and maybe
more importantly, by applying the new theory to various models, as it
was done with its bulk counterpart, a thorough exploration of the
phenomenology of confined glassforming systems would become possible,
potentially allowing to disentangle the different physical effects
which interplay in these systems.

It is the aim of this Letter to provide such an extension of the MCT
for a particular class of confined systems, the so-called
``quenched-annealed'' (QA) binary mixtures. In these systems, first
introduced by Madden and Glandt \cite{MG1988}, the fluid molecules
equilibrate in a matrix of particles frozen in a disordered
configuration sampled from a given probability distribution. The
models studied in Refs.~\cite{galpelrov02el,kim03el,chajagyet04pre}
belong to this class of systems, as does the Lorentz model, which
corresponds to a zero fluid density limit. Besides the proposed theory
will borrow ideas from the mode coupling approaches to the
diffusion-localization transition in this model
\cite{gotleuyip81pra,leu83pra,sza04el}.

The theory is derived using the projection operator method described
in Ref.~\cite{leshouches}. In the present problem, the inner product
of two arbitrary dynamical variables $A$ and $B$ is given by
$\overline{\langle A B^* \rangle}$, where $^*$ denotes complex
conjugation, $\langle \cdots \rangle$ a thermal average \emph{taken
for a given realization of the matrix} and $\overline{\cdots}$ a
\emph{subsequent} average over the matrix realizations. We thus
consider a fluid made of $N_f$ particles of mass $m$, adsorbed in a
homogeneous disordered matrix consisting of $N_m$ immobile sites. The
system has volume $V$, hence the fluid and matrix densities are
respectively $n_f=N_f/V$ and $n_m=N_m/V$. As in the bulk MCT, the
dynamical variables of interest are the Fourier components of the
microscopic fluid density, $\rho^f_\mathbf{q}(t)=\sum_{j=1}^{N_f} e^{i
\mathbf{q} \mathbf{r}_j(t)}$, where $\mathbf{q}$ denotes the
wavevector and $\mathbf{r}_j(t)$ is the position of the fluid particle
$j$ at time $t$. But, before proceeding with the dynamical theory, one
has to take care of certain peculiarities of the statics of QA
systems. Indeed, because of the presence of the quenched component,
for a given matrix realization, the translational invariance of the
system is broken. This implies that, at variance with bulk fluids,
time-persistent density fluctuations exist at equilibrium,
i.e., $\langle \rho^f_\mathbf{q} \rangle \neq 0$. This fact is well
known from the derivation, using the replica method, of the
Ornstein-Zernike (OZ) equations describing this type of systems
\cite{G1992,RTS1994,MD1994}, where it leads to the splitting of the
total and direct correlation functions of the fluid, respectively
$h^{ff}(r)$ and $c^{ff}(r)$, into two contributions, connected
[$h^{c}(r)$ and $c^{c}(r)$] and blocked or disconnected [$h^{b}(r)$
and $c^{b}(r)$]. A similar splitting occurs for the structure factor
of the fluid $S^{ff}_q= \overline{\langle \rho^f_\mathbf{q}
\rho^{f}_\mathbf{-q} \rangle} / N_f =1+n_f \hat{h}^{ff}_q$, leading to
$S^{ff}_q = S^{c}_q + S^{b}_q$ with $S^{c}_q = \overline{\langle(
\rho^f_\mathbf{q} - \langle\rho^f_\mathbf{q}\rangle)
(\rho^{f}_\mathbf{-q}-\langle\rho^{f}_\mathbf{-q} \rangle)\rangle} /
N_f=1+n_f \hat{h}^{c}_q$ and $S^{b}_q =
\overline{\langle\rho^f_\mathbf{q}\rangle \langle\rho^{f}_\mathbf{-q}
\rangle}/N_f=n_f \hat{h}^{b}_q$, where $\hat{f}_q$ denotes the Fourier
transform of $f(r)$. For future reference, we define the matrix-matrix
and fluid-matrix structure factors and total correlation functions as
well, which are given by $S^{mm}_q = \overline{\langle
\rho^m_\mathbf{q} \rho^{m}_\mathbf{-q}\rangle}/ N_m=1+n_m
\hat{h}^{mm}_q$ and $S^{fm}_q = \overline{\langle \rho^f_\mathbf{q}
\rho^{m}_\mathbf{-q} \rangle} / \sqrt{N_f N_m}= \sqrt{n_f n_m}
\hat{h}^{fm}_q$; $\rho^m_\mathbf{q}=\sum_{j=1}^{N_m} e^{i \mathbf{q}
\mathbf{s}_j}$, where $\mathbf{s}_j$ is the fixed position of the
matrix particle $j$, is the $\mathbf{q}$ Fourier component of the
quenched microscopic matrix density.

It thus follows that, if one is only interested in the relaxing part
of the fluid density fluctuations, one has to consider the dynamical
variable $\delta\rho^f_\mathbf{q}(t)=\rho^f_\mathbf{q}(t)-\langle
\rho^f_\mathbf{q} \rangle$ rather than $\rho^f_\mathbf{q}(t)$ itself.
Using the standard method, a generalized Langevin equation for the
time evolution of the normalized autocorrelation function of the
connected density fluctuations $\phi_q(t)= \overline{\langle
\delta\rho^f_\mathbf{q}(t) \delta\rho^f_\mathbf{-q}\rangle}/ (N_f
S^{c}_q)$ can then be derived, which is formally identical to the
equation for bulk fluids, i.e.,
\begin{equation}\label{langcoll}
\ddot{\phi}_{q}+\Omega_{q}^2 \phi_{q}+ \Omega_{q}^2 \int_0^t d\tau
M_q(t-\tau) \dot{\phi}_{q}(\tau)=0,
\end{equation}
with $\Omega_{q}^2=q^2 k_B T /(m S^{c}_q)$, where $T$ is the
temperature and $k_B$ the Boltzmann constant. The memory function is
given by $\Omega_{q}^2 M_q(t)= \overline{\langle
R_\mathbf{q}(t)\,R_\mathbf{-q}\rangle}/(N_f m k_B T)$, where
$R_\mathbf{q}(t) =\exp[i (1-\mathscr{P}) \mathscr{L} (1-\mathscr{P})
t] i (1-\mathscr{P}) \mathscr{L} g^f_\mathbf{q}$ is the projected
random force obtained from the longitudinal fluid momentum density
fluctuation $g^f_\mathbf{q}(t)$. $\mathscr{L}$ is the Liouville
operator of the system and $\mathscr{P}$ is the projector onto the
subspace of dynamical variables spanned by $\delta\rho^f_\mathbf{q}$
and $g^f_\mathbf{q}$.

We now obtain the slow decaying portion of the memory kernel with a
mode coupling approach, assuming that the slow decay of the projected
random force autocorrelation function is dominated by couplings to two
types of quadratic dynamical variables,
$B^{(1)}_{\mathbf{q,k}}=\delta\rho^f_\mathbf{k}
\delta\rho^f_\mathbf{q-k}$, in analogy with the bulk MCT, and
$B^{(2)}_{\mathbf{q,k}}=\delta\rho^f_\mathbf{k} \rho^m_\mathbf{q-k}$,
following previous work on the Lorentz gas
\cite{gotleuyip81pra,leu83pra,sza04el}. Defining the projector
$\mathcal{P}$ on the subspace spanned by the $B_{\mathbf{q,k}}$'s, the
mode coupling part of the memory function is then expressed as
$\Omega_{q}^2 M^{(\text{MC})}_q(t)=\overline{\langle
\mathcal{P}R_\mathbf{q}(t)\,\mathcal{P}R_\mathbf{-q} \rangle}/(N_f m
k_B T)$.  To complete the calculation of $M^{(\text{MC})}_q$, two
steps remain. First, a factorization approximation is required to
express four-point density correlations into products of two-point
density correlations.  Following the usual mode coupling prescription
then leads to (we note $\mathscr{Q}=1-\mathscr{P}$)
\begin{gather*}
\overline{\langle (e^{i\mathscr{QLQ}t} \delta\rho^f_\mathbf{k}
\delta\rho^f_\mathbf{q-k} ) \delta\rho^f_\mathbf{-k'}
\delta\rho^f_\mathbf{-q+k'} \rangle} \simeq (\delta_{\mathbf{k,k'}}
+ \delta_{\mathbf{k,q-k'}}) N_f^2 S^{c}_k S^{c}_{|\mathbf{q-k}|}
\phi_{k}(t) \phi_{|\mathbf{q-k}|}(t),\\
\overline{ \langle (e^{i\mathscr{QLQ}t} \delta\rho^f_\mathbf{k}
\rho^m_\mathbf{q-k})\delta\rho^f_\mathbf{-k'} \rho^m_\mathbf{-q+k'}
\rangle} \simeq \delta_{\mathbf{k,k'}} N_f N_m S^{c}_k
S^{mm}_{|\mathbf{q-k}|}\phi_{k}(t),\\ 
\overline{\langle ( e^{i\mathscr{QLQ}t} \delta\rho^f_\mathbf{k}
\delta\rho^f_\mathbf{q-k} ) \delta\rho^f_\mathbf{-k'}
\rho^m_\mathbf{-q+k'} \rangle} \simeq 0.
\end{gather*}
A crucial point here is that, since the matrix is quenched,
$\rho^m_\mathbf{q}$ shows no thermal fluctuations: $\overline{ \langle
\delta\rho^f_\mathbf{q}(t) \rho^m_\mathbf{-q}\rangle}$ is thus
identically zero [remember that $\delta\rho^f_\mathbf{q}(t) =
\rho^f_\mathbf{q}(t)-\langle \rho^f_\mathbf{q} \rangle$].

Second, one needs to calculate $\overline{\langle R_\mathbf{q}
B^{(l)}_{\mathbf{-q,-k}} \rangle} = \overline{\langle
i\mathscr{L}g^f_\mathbf{q}
B^{(l)}_{\mathbf{-q,-k}}\rangle}-\overline{\langle
i\mathscr{PL}g^f_\mathbf{q} B^{(l)}_{\mathbf{-q,-k}}\rangle}$. The
first term is readily handled by application of the Yvon theorem, just
as in the bulk MCT. The second one is far more delicate, since it
involves three-point connected static correlations of the QA
system. Usually, such terms are estimated using the so-called
convolution approximation \cite{JacFee62rmp} which leads to remarkable
simplifications in the resulting mode coupling equations.  An
extension of the convolution approximation to QA systems has thus been
developed which gives
\begin{align*}
\overline{\langle \delta\rho^f_{\mathbf{q}}
\delta\rho^{f}_{\mathbf{-k}} \delta\rho^{f}_{\mathbf{-q+k}} \rangle}&=
N_f S^{c}_q S^{c}_k S^{c}_{|\mathbf{q-k}|},\\ \overline{\langle
\delta\rho^f_{\mathbf{q}} \delta\rho^{f}_{\mathbf{-k}}
\rho^{m}_{\mathbf{-q+k}}\rangle}&=  \sqrt{N_f N_m} S^{c}_q S^{c}_k
S^{fm}_{|\mathbf{q-k}|}.
\end{align*}

Eventually, assuming that the contributions to the memory kernel not
included in $M^{(\text{MC})}_q$ can be replaced by a white noise term
$\Gamma_q \delta(t)$, we obtain the mode coupling equations for the
collective dynamics of a QA mixture, Eq.~\eqref{langcoll} with
$M_q(t)=\Gamma_q \delta(t) + M^{(\text{MC})}_q(t)$ and
\begin{subequations}\label{kerncoll}
\begin{gather}
M^{(\text{MC})}_q(t)=\frac{1}{V} \sum_{\mathbf{k}}
V^{(2)}_{\mathbf{q},\mathbf{k}} \phi_{k}(t) \phi_{|\mathbf{q-k}|}(t) +
V^{(1)}_{\mathbf{q},\mathbf{k}} \phi_{k}(t),\\
V^{(2)}_{\mathbf{q},\mathbf{k}} = \frac{1}{2} n_f S^{c}_q
\left[\frac{\mathbf{q}\cdot\mathbf{k}}{q^2} \hat{c}^{c}_k +
\frac{\mathbf{q}\cdot(\mathbf{q-k})}{q^2}
\hat{c}^{c}_{|\mathbf{q-k}|}\right]^2 S^{c}_k
S^{c}_{|\mathbf{q-k}|},\\
V^{(1)}_{\mathbf{q},\mathbf{k}} = n_m S^{c}_q
\left[\frac{\mathbf{q}\cdot(\mathbf{q-k})}{q^2}+n_f
\frac{\mathbf{q}\cdot\mathbf{k}}{q^2} \hat{c}^{c}_k \right]^2
\frac{(\hat{h}^{fm}_{|\mathbf{q-k}|})^2}{S^{mm}_{|\mathbf{q-k}|}}
S^{c}_k,
\end{gather}
\end{subequations}
where the replica OZ equations were used to introduce the relevant
direct correlation functions \cite{G1992,RTS1994}.

The same procedure can be applied to the dynamics of a tagged particle
and the corresponding equations will be reported in a forthcoming
paper.

Equations \eqref{langcoll} and \eqref{kerncoll} form the proposed MCT
for QA mixtures. Like in the bulk, they involve static quantities
only, and, more crucially, they retain the mathematical structure of
the typical mode coupling equations which have been extensively
studied in Ref.~\cite{leshouches}. Thus, all the known properties of
the solutions of MCT equations, in particular their universal
behaviors, apply \textit{a priori} unchanged to QA mixtures. This
means that, in principle, the analysis performed in
Ref.~\cite{galpelrov02el} is as legitimate as all the analogous ones
done on bulk systems.

As one would expect, the present theory integrates the previously
known mode coupling theories as limiting cases: In the limit of
vanishing matrix density, the bulk MCT \cite{bengotsjo84jpc} is
recovered, while in the limit of vanishing fluid density, the MCT
equations for the Lorentz gas \cite{sza04el} are obtained. Since both
limits show ergodicity-breaking transitions (ideal glass transitions
in the first case, diffusion-localization transitions in the second),
the present theory, which ``interpolates'' between them, is bound to
display such phenomena.

To illustrate this point and as a first demonstration of the
potentialities of the theory, we have computed the dynamical phase
diagram of a simple QA system (the models of
Refs.~\cite{galpelrov02el,kim03el} would be quite complex for a
preliminary study). This is the one studied in
Ref.~\cite{chajagyet04pre}, which consists of a fluid of hard spheres
confined in a matrix of hard spheres frozen in an equilibrium
configuration. Both the fluid and matrix particles have diameter
$\sigma$, and the system is characterized by two volume fractions
$\phi_f=\pi n_f\sigma^3/6$ and $\phi_m=\pi n_m\sigma^3/6$. The
Percus-Yevick approximation \cite{G1992,MLW96JCP} is used to compute
the required structural quantities.  Since we are confronted to
basically the same equations as in bulk systems, the numerical
procedures to track ergodicity-breaking transitions signalled by the
appearance of a nonzero infinite time limit to $\phi_{q}(t)$ do not
differ from those used there. We have applied the method which is
described in Ref.~\cite{frafucgotmaysin97pre}.

\begin{figure}
\includegraphics{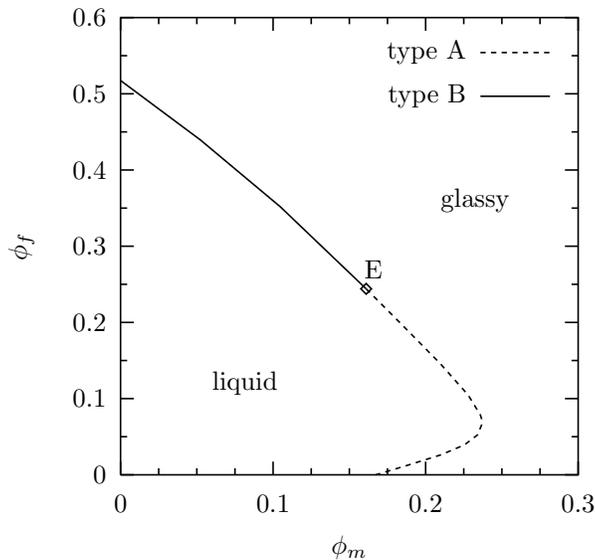}
\caption{\label{fig1} Dynamical phase diagram of a hard sphere fluid
confined in a matrix of identical hard spheres frozen in an equilibrium
configuration. $\phi_f$ and $\phi_m$ denote respectively the fluid and
matrix compacities. Point E is the common endpoint of the
type A and type B transition lines.}
\end{figure}

The corresponding dynamical phase diagram is reported in
Fig.~\ref{fig1}.  It consists of two transition lines. On the one
hand, starting from the bulk fluid ($\phi_m=0$) transition point and
increasing $\phi_m$, one follows a line of discontinuous or type B
transitions, where $f_q=\lim_{t\to\infty} \phi_{q}(t)$ jumps
discontinuously from zero to a nonzero value when moving from the
ergodic liquid phase to the nonergodic glassy phase. Along this line,
as $\phi_m$ is increased, the amplitude of the jump decreases to zero
and the exponent parameter $\lambda$, which determines many properties
of the solutions of the mode coupling equations
\cite{leshouches,gotsjo92rpp,got99jpcm}, increases from its bulk value
(of about $.73$) to one, its largest allowed value. On the other hand,
moving away from the diffusion-localization ($\phi_f=0$) transition
point by increasing $\phi_f$, one follows a line of continuous or type
A transitions, from which $f_q$ grows continuously from zero when the
system enters in the glassy domain. Here, as $\phi_f$ is increased,
$\lambda$ grows continuously from zero to one, the allowed interval
for type A transitions.

Both lines connect smoothly at a common endpoint E, where $\lambda=1$
for both. Point E corresponds to a degenerate A$_3$ singularity, a
generic type of topologically stable singularities already known from
the so-called $F_{12}$ model \cite{f12}. To our knowledge, this widely
studied one equation toy model had never found any physical
realization until now. This result has important physical
implications, since, in the vicinity of such a higher-order
singularity, the dynamics are known to display very specific features,
most significantly logarithmic decay laws and subdiffusive behaviors
\cite{leshouches,highorder,gotspe02pre}.

Beside this specific bifurcation scenario, and formally not related to
it, another remarkable prediction of the present theory lies in the
shape of the phase diagram. Indeed, starting from the zero fluid
density limit and increasing $\phi_f$, the matrix density at which the
system becomes frozen first increases, reaches a maximum and then
decreases until the bulk limit is reached.  The last behavior can be
easily understood from simple free volume arguments: Because of the
volume excluded by the matrix particles, the larger the matrix density
is, the smaller the fluid density has to be for structural arrest to
occur. The first regime however is quite unexpected and might be due
to a partial relaxation of the dynamical correlations which lead to
the localization of a single particle moving in the porous medium by
the introduction of collective density fluctuations at a finite but
small fluid density. Overall, this nonmonotonic behavior of $\phi_m$
at the dynamical transition leads to reentrant type A transitions,
i.e., for a given matrix density, ergodicity can be broken both by an
increase or a decrease of the fluid density.

All these predictions (bifurcation scenario, logarithmic decay laws,
shape of the transition lines, evolutions of $f_q$ and $\lambda$ along
these lines) can be tested by computer simulations to judge of the
validity of the present theory. Unfortunately, the focus of the work
of Ref.~\cite{chajagyet04pre} was not on a putative mode coupling
scenario and thus only indirect and not so convincing comparisons in
favor of the theory can be made. For instance, these authors found
that at the lowest investigated fluid density ($\phi_f=0.05$), the
dependence of the diffusion coefficient on $\phi_m$ was different from
the one found at higher fluid densities. This might be a signature of
the nonmonotonicity of the transition line in this low fluid density
regime. Another of their observations, made in Ref.~\cite{kim03el} as
well and not so surprising, is that the inclusion of matrix particles
slows down the dynamics more efficiently than the inclusion of the
same amount of fluid particles. Here, this is reflected in the fact
that the total compacity $\phi_\text{tot}=\phi_f+\phi_m$ at the
transition is a decreasing function of $\phi_m$ in the top part of the
phase diagram of Fig.~\ref{fig1}. Thus, at a transition point, the
corresponding value of $\phi_\text{tot}$ being held fixed, an increase
of $\phi_f$ at the expense of $\phi_m$ leads to an ergodic system,
while the reverse change drives the system deeper into the nonergodic
domain. This looks encouraging, but clearly more simulation work is
needed.

In summary, we have developed an extension of the MCT to the QA
mixture model of confined fluids. The corresponding equations turn out
to be similar to those of the MCT for the bulk, so that all the
applications of the theory which have been conceivable for the bulk,
like tests of its universal predictions or quantitative comparisons
with computer simulations for simple models, are transposable to the
present class of systems.  The calculation of the dynamical phase
diagram of a simple system shows that new and complex bifurcation
scenarios can be predicted and that a rich phenomenology could be
unveiled by a systematic study of models of increasing
complexity. Such a work is under way.

This of course does not exhaust the question of a general
mode coupling description of confined glassforming fluids. Indeed, the
QA mixture has the simplifying feature that it corresponds to a
statistically homogeneous confinement, while many studies have been
done for slit, cylindrical or spherical geometries of the confining
medium. The present development should nevertheless represent a
valuable means to improve our general understanding of the slow
dynamics of confined glassforming liquids.

\acknowledgments It is a pleasure to thank W. G{\"o}tze for useful
comments and G. Tarjus for fruitful discussions and, together with
M.L. Rosinberg and E. Kierlik, for an earlier collaboration on the
theory of QA mixtures which made this work possible.


\begin{thebibliography}{99}
\bibitem{leu84pra} E. Leutheusser, Phys. Rev. A \textbf{29}, 2765
(1984).
\bibitem{bengotsjo84jpc} U. Bengtzelius, W. G{\"o}tze, and
A.~Sj{\"o}lander, J. Phys. C \textbf{17}, 5915 (1984).
\bibitem{leshouches} W. G{\"o}tze, in \textit{Liquids, freezing and
glass transition}, edited by J.-P.~Hansen, D. Levesque, and
J. Zinn-Justin (North Holland, Amsterdam, 1991), pp.~287-503.
\bibitem{gotsjo92rpp} W. G{\"o}tze and L. Sj{\"o}gren,
Rep. Prog. Phys. \textbf{55}, 241 (1992).
\bibitem{got99jpcm} W. G{\"o}tze, J. Phys.: Condens. Matter
\textbf{11}, A1 (1999).
\bibitem{highorder} A recent example is the study of short-ranged
attractive colloids; see F. Sciortino, P. Tartaglia, and
E. Zaccarelli, Phys. Rev. Lett. \textbf{91}, 268301 (2003), and
references therein.
\bibitem{proceedings} See for instance \textit{Proceedings of the
International Workshop on Dynamics in Confinement}, J. Phys. (Paris)
IV \textbf{10}, Pr7-203 (2000), and \textit{Proceedings of Second
International Workshop on Dynamics in Confinement}, Eur. Phys. J. E
\textbf{12}, 3-204 (2003).
\bibitem{heterogeneities} H. Sillescu, J. Non-Cryst. Solids
\textbf{243}, 81 (1999); M.D. Ediger,
Annu. Rev. Phys. Chem. \textbf{51}, 99 (2000); R. Richert, J. Phys.:
Condens. Matter \textbf{14}, R703 (2002).
\bibitem{galrovspo00prl} P. Gallo, M. Rovere, and E. Spohr,
Phys. Rev. Lett. \textbf{85}, 4317 (2000);
J. Chem. Phys. \textbf{113}, 11324 (2000).
\bibitem{galpelrov02el} P. Gallo, R. Pellarin, and M. Rovere,
Europhys. Lett. \textbf{57}, 212 (2002); Phys. Rev. E \textbf{67},
041202 (2003); \textit{ibid.} \textbf{68}, 061209 (2003).
\bibitem{schkolbin04jpcb} P. Scheidler, W. Kob, and K. Binder,
J. Phys. Chem. B \textbf{108}, 6673 (2004).
\bibitem{MG1988} W.G. Madden and E.D. Glandt,
J. Stat. Phys. \textbf{51}, 537 (1988); W.G. Madden,
J. Chem. Phys. \textbf{96}, 5422 (1992).
\bibitem{kim03el} K. Kim, Europhys. Lett. \textbf{61}, 790 (2003).
\bibitem{chajagyet04pre} R. Chang, K. Jagannathan, and A. Yethiraj,
Phys. Rev. E \textbf{69}, 051101 (2004).
\bibitem{gotleuyip81pra} W. G{\"o}tze, E. Leutheusser, and S. Yip,
Phys. Rev. A \textbf{23}, 2634 (1981).
\bibitem{leu83pra} E. Leutheusser, Phys. Rev. A \textbf{28}, 2510
(1983).
\bibitem{sza04el} G. Szamel, Europhys. Lett. \textbf{65}, 498 (2004).
\bibitem{G1992} J.A. Given and G. Stell, J. Chem. Phys. \textbf{97},
4573 (1992); E. Lomba, J.A. Given, G. Stell, J.J. Weis, and
D. Levesque, Phys. Rev. E \textbf{48}, 233 (1993); J.A. Given and
G. Stell, Physica A \textbf{209}, 495 (1994).
\bibitem{RTS1994} M.L. Rosinberg, G. Tarjus, and G. Stell,
J. Chem. Phys. \textbf{100}, 5172 (1994).
\bibitem{MD1994} G. I. Menon and C. Dasgupta,
Phys. Rev. Lett. \textbf{73}, 1023 (1994).
\bibitem{JacFee62rmp} H.W. Jackson and E. Feenberg,
Rev. Mod. Phys. \textbf{34}, 686 (1962).
\bibitem{MLW96JCP} A. Meroni, D. Levesque, and J.J. Weis,
J. Chem. Phys. \textbf{105}, 1101 (1996).
\bibitem{frafucgotmaysin97pre} T. Franosch, M. Fuchs, W. G{\"o}tze,
M.R. Mayr, and A.P. Singh, Phys. Rev. E \textbf{55}, 7153 (1997).
\bibitem{f12} W. G{\"o}tze, Z. Phys. B \textbf{56}, 139 (1984).
\bibitem{gotspe02pre} W. G\"otze and M. Sperl, Phys. Rev. E
\textbf{66}, 011405 (2002); M. Sperl, \textit{ibid.} \textbf{68},
031405 (2003); W. G\"otze and M. Sperl, J. Phys.: Condens. Matter
\textbf{16}, S4807 (2004).
\end{thebibliography}
\end{document}